\documentclass[letterpaper, 10 pt, conference]{ieeeconf}  

\usepackage{amsmath,amsfonts}

\usepackage{array}
\usepackage[caption=false,font=normalsize,labelfont=sf,textfont=sf]{subfig}
\usepackage{textcomp}
\usepackage{stfloats}
\usepackage{url}
\usepackage{verbatim}
\usepackage{graphicx}
\usepackage{cite}
\usepackage[utf8]{inputenc}
\usepackage[letterpaper,top=2cm,bottom=2cm,left=2cm,right=2cm,marginparwidth=1.75cm]{geometry}
\usepackage[T1]{fontenc} 
\usepackage[linesnumbered,ruled]{algorithm2e}

\usepackage{algpseudocode}
\usepackage{graphicx}    
\usepackage{caption}    
\usepackage{amsmath}    
\usepackage{amssymb}    
\usepackage{amsthm}
\usepackage{indentfirst} 

\usepackage{color}      
\twocolumn              

\IEEEoverridecommandlockouts                              

\title{\LARGE \bf
Hierarchical Climate Control Strategy for Electric Vehicles with Door-Opening Consideration
}

\author{Sanghyeon Nam$^{1}$, Hyejin Lee$^{1}$, Youngki Kim$^{2}$, Kyoung hyun Kwak$^{2}$ and Kyoungseok Han$^{1}$
\thanks{*This research was supported by Basic Science Research Program through the National Research Foundation of Korea (NRF) funded by the Ministry of Education(2021R1A6A1A03043144); in part by the NRF grant, funded by the Korean government (MSIT) (NRF-2021R1C1C1003464); in part by the Technology Innovation Program ('20021926', 'Development of Eco-friendly Vehicle Tuning Supported Open Platform using Design and Verification Technology for Carbon Neutrality.}
\thanks{$^{1}$School of Mechanical Engineering, Kyunpook National University, Daegu 41566, Korea
        {\tt\small (nju1245, gpwls2014, kyoungsh) @knu.ac.kr}}%
\thanks{$^{2}$The Department of Mechanical Engineering, Uiversity of Michigan-Dearborn, Dearborn, MI 48128, USA
        {\tt\small (youngki, khkwak)@umich.edu}}%
}

\begin{document}

\maketitle
\thispagestyle{empty}
\pagestyle{empty}

\begin{abstract}

This study proposes a novel climate control strategy
for electric vehicles (EVs) by addressing door-opening
interruptions, an overlooked aspect in EV thermal management.
We create and validate an EV simulation model that
incorporates door-opening scenarios. Three controllers are
compared using the simulation model: (i) a hierarchical nonlinear
model predictive control (NMPC) with a unique coolant
dividing layer and a component for cabin air inflow regulation
based on door-opening signals; (ii) a single MPC controller;
and (iii) a rule-based controller. The hierarchical controller
outperforms, reducing door-opening temperature drops by
46.96\% and 51.33\% compared to single layer MPC and
rule-based methods in the relevant section. Additionally, our
strategy minimizes the maximum temperature gaps between
the sections during recovery by 86.4\% and 78.7\%, surpassing
single layer MPC and rule-based approaches, respectively.
We believe that this result opens up future possibilities for
incorporating the thermal comfort of passengers across all
sections within the vehicle.

\end{abstract}

\section{INTRODUCTION}

Growing environmental concerns and stricter emissions regulations are driving the shift to electric vehicles (EVs), especially in taxis. However, cold climates pose challenges due to the lack of an additional heat source. Addressing this, a study highlights the importance of a precise real-time EV energy consumption model \cite{steinstraeter2021effect}, with an average error of 5.9\% compared to empirical data. The analysis also examines the impact of reduced power capacity on battery efficiency, utilizing models from \cite{pesaran2013addressing} and \cite{qian2010temperature}.

Studies utilizing refined EV driving models focus on mitigating range reduction through advanced thermal management control strategies. \cite{glos2021non} employs nonlinear model predictive control (NMPC) to regulate cabin temperature and air quality, addressing waste heat limitations. \cite{kwak2023thermal} introduces an MPC-based eco-climate control system for EV HVAC, surpassing rule-based controllers in energy efficiency and thermal comfort. Meanwhile, \cite{hajidavalloo2023nmpc} presents an NMPC-based integrated thermal management strategy for EVs, optimizing driving range and cabin comfort in real time, outperforming rule-based controls.

In addition, citing notable works such as \cite{lahlou2018dynamic, konda2022energy, amini2019cabin, hu2023robust, zhao2021two, ghalkhani2022review}, where these contributions feature advanced algorithms optimizing energy consumption under specific driving conditions. Despite efforts, a research gap persists, especially in addressing disruptions from routine actions like door or window openings, and no significant studies on this matter have been reported to date. As we envision driving an electric vehicle during the cold winter season, significant temperature drops across the entire cabin, especially in areas where doors are opened, become apparent. However, typical studies overlook individual sections, impeding the ability to address interruptions at a sectional level. The growing prevalence of electric vehicles, including taxis, underscores the urgent need for thermal control strategies at a sectional level that account for these routine actions.

The research \cite{luchini2020model} focuses on temperature control for a food supplier vehicle, optimizing it through an MPC-based approach that addresses the effects of door-opening. Notably, however, it excludes scenarios involving considerations related to passengers, and does not specifically focus on EV systems. Other studies like \cite{ma2015influence}, \cite{hasanuzzaman2008investigation}, and \cite{marr2012influence} explore door-opening heat loss in various contexts, but not within the specific realm of transportation systems.

Considering the scenario of easily occurring door-opening interruptions, this paper outlines the following key contributions: 

(i) We develop a practical simulation model designed for capturing the underlying dynamics of the electric powertrain of EV during operation under specific driving cycles, with a specific emphasis on the HVAC system. We validate the model through real car experiments. Moreover, our model includes door-opening scenarios, which is a novel addition not explored in previous studies, thereby enhancing its practical applicability.

(ii) We introduce a hierarchical control strategy utilizing NMPC with two layers to optimize energy efficiency and proficiently address interruptions from door-opening events. The first component divides the coolant flow through a 3-way valve, while the second, a conditional layer, directs inflowing air into the cabin. This involves deploying a conditional MPC controller specifically activated during door-opening events, with the purpose of reducing the computational burden when door-opening does not occur. This approach enables us to minimize temperature recovery time by precisely regulating air flows.

(iii) Simulation results of our NMPC-based thermal management strategy, alongside benchmark control strategies like rule-based controllers, showcase the effectiveness of our hierarchical control approach, particularly in effectively handling door-opening scenarios critical for maintaining the desired temperature in the cabin.

The remainder of the paper is organized as follows: Section II introduces the ITMS and validated door-opening model. Section III details the hierarchical heating control strategy, emphasizing the primary contribution. Section IV presents case studies comparing our approach with baselines. Section V concludes, summarizing key findings and contributions.

\begin{figure*}[!t]
\centering
\vspace{3mm}
\includegraphics[width=160mm]{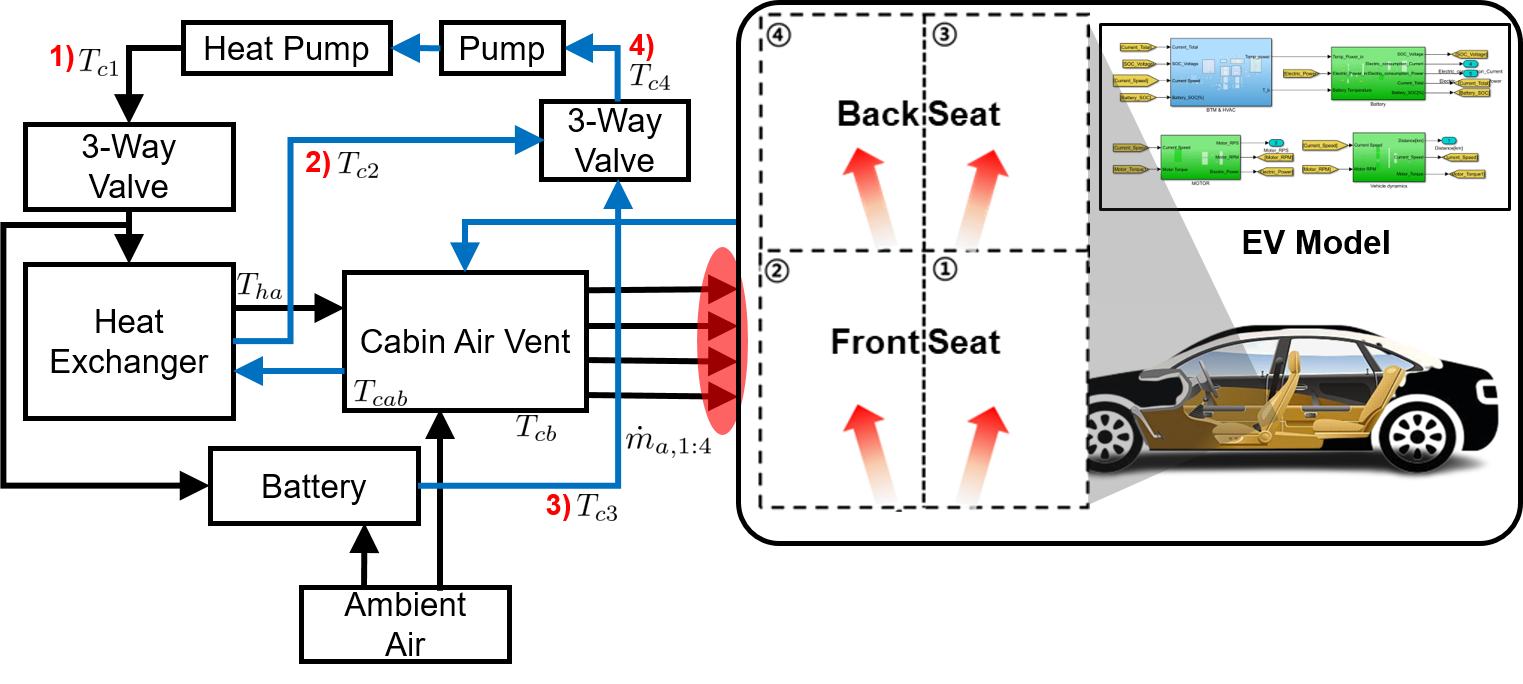}
\caption{The illustration of the integrated thermal management system in this paper.}
\label{Fig_Big1}
\end{figure*}

\section{MODELING}

In this section, we introduce our door-opening considered cabin model in detail. The longitudinal dynamics of the vehicle are employed based on the reference \cite{rajamani2011vehicle}. 
Given space constraints, a brief overview of the coolant cycle is presented here. For further details, refer to our prior work \cite{nam2024design}, where we developed an EV model with a 150 kW traction motor, a 64 kWh battery pack, and an ITMS, incorporating thermal interaction models.

\subsection{Cabin Model}

In Fig.~\ref{Fig_Big1}, the coolant cycle is segmented into four parts: 1) after the heat pump heat exchanger, 2) after the cabin-side heat exchanger, 3) after interaction with the battery, and 4) the combined temperature of 2) and 3). This paper uses the notations $T_{c1}$, $T_{c2}$, $T_{c3}$, and $T_{c4}$ to represent the coolant temperatures of these relevant phases.
In consideration of limited space, as highlighted earlier, we opt to omit the detailed temporal evolution formulations for the coolant phases in this paper. This choice is made to center our attention on the main contribution of this paper, which is the cabin model addressing door-opening interruptions.

Firstly, the formulation for the standard cabin model formulations, excluding considerations for door-opening, encompasses its interaction with the cabin body $T_{cb}$ and cold ambient air $T_{amb}$ are expressed as follows:
\begin{align}
\label{1}
\small
& \frac{dT_{cab}}{dt} = \nonumber \\
& \frac{\alpha_{cab}A_{cb}(T_{cb}-T_{cab})+\dot{m}_{a}c_{a}(T_{ha}-T_{cab})+\dot{Q}_{occ}}{m_{a}c_{a}}, \\
& \frac{dT_{cb}}{dt} = \nonumber \\
& \frac{\alpha_{cb}A_{cb}(T_{cab}-T_{cb})+\alpha_{cb}A_{cb}(T_{amb}-T_{cb})+\dot{Q}_{sol}}{m_{cb}c_{cb}}
\end{align}
here, $\alpha_{cab}$ and $\alpha_{cb}$ represent the thermal interaction coefficients of the cabin and cabin body, respectively, while $A_{cb}$ denotes the surface area of the cabin body. $\dot{Q}_{occ}$ and $\dot{Q}_{sol}$ denote heat generation by the occupants in the cabin and the transmitted heat from solar radiation, respectively. Additionally, $m_{a}$, $m_{cb}$, $c_{a}$, and $c_{cb}$ stand for the mass of air, cabin body, specific heat capacity of air, and specific heat capacity of the cabin body, respectively.
The variable $\dot{m}_{a}$ represents the air inflow into the cabin, serving as a decision variable within this model.
 
The $T_{ha}$ stands for the heated air which flows into the cabin after thermal interaction with coolant at the heat exchanger and its temporal evolution formulation is as follow:
\begin{equation}
\label{2}
    \frac{dT_{ha}}{dt} = \frac{\dot{m}_{a}c_{a}(T_{cab}-T_{ha})+\Gamma_{HX}(T_{c2}-T_{ha})}{m_{a}c_{a}}
\end{equation}
here, $\Gamma_{HX}$ stands for the heat transfer coefficient between the inlet air and the coolant in $W/K$.

\begin{figure}[h]
    \centering
    \vspace{4mm}
    \includegraphics[width =82mm]{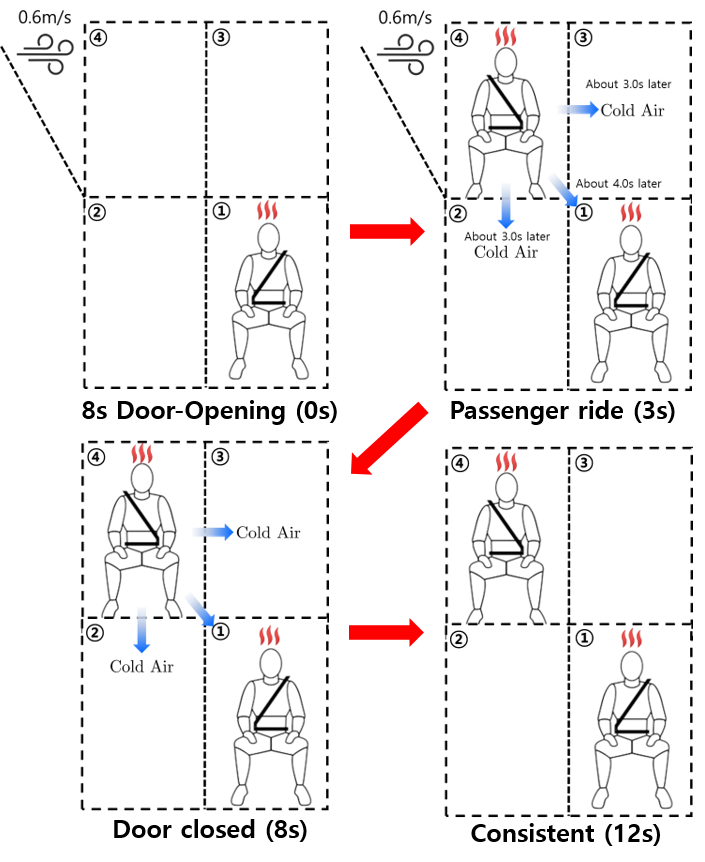}
    \caption{The figure depicting cold air propagation in the four distinct cabin sections model due to door-opening in Section 4.}
    \label{Cabin Section}
\end{figure}

Based on these formulations, we construct separated cabin sections, divided into four distinct parts, to effectively elucidate thermal interactions between sections during door-opening events. The door-opening model, $f_{\text{open}}$, constructed in this paper, is based on a detailed analysis from a formal study \cite{nagano2016effect} exploring the thermal impact of door-opening in cold climates. 

Figure~\ref{Cabin Section} represents the four distinguished sections of the cabin. We assumed that the door-opening occurs in Section 4 in this study. When door-opening occurs, cold air flows into the cabin and that chill propagates to the other sections after 3 seconds, in general. The formulations for the door-opening cabin models in each section, denoted as $T_{s1:4} \gets T_{s1}, T_{s2}, T_{s3}, T_{s4},$ are as follows:
\begin{align}
\label{DoorEQ}
\small
    & \frac{dT_{s1:4}}{dt} =  \nonumber \\
    &\frac{\alpha_{cb}A_{cb,1:4}(T_{cb}-T_{s1:4})+\dot{m}_{a,1:4}c_{a}(T_{ha}-T_{s1:4})+\dot{Q}_{add}}{m_{s1:4}c_{a}}.
\end{align}
In this equation, $\dot{Q}_{add}$ represents the aggregate of additional heat generation and losses for each section, including contributions from occupants and losses due to door-opening. Crucially, this value is a time-varying variable, dependent on time and variable across sections. The variables $A_{cb,1:4}$ and $m_{s1:4}$ represent the surface area of the cabin body and the mass of air for each section, respectively. The value $\dot{m}_{a,1:4}$ represent the air inflow for each section. Further details regarding this variable will be elaborated upon in Section III.

\subsection{Model Validation}

\begin{figure}[t]
    \centering
    \vspace{4mm}
    \includegraphics[width =78mm]{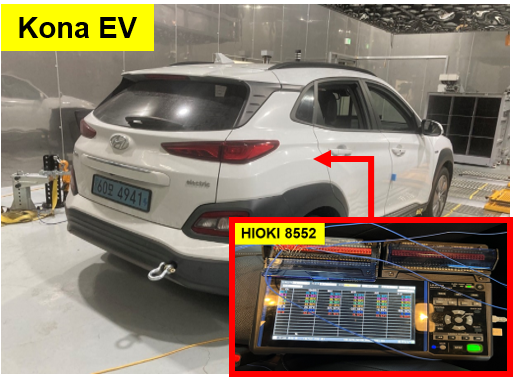}
    \caption{KONA EV and HIOKI thermometer utilized for the experiment}
    \label{Hioki}
\end{figure}

In this subsection, we validated our model with a real car experiment at KATECH, Daegu, South Korea, using a KONA EV by HYUNDAI, mirroring our ITMS model specifications. The experiment used a HIOKI 8552 voltage/temp unit, capable of measuring 30 channels at a 20ms sampling time. Conducted in an environment with -7°C ambient temperature, the experiment targeted a cabin temperature of precisely 23°C, aligning with simulation conditions. Figure~\ref{Hioki} displays the vehicle and thermometer used in this validation experiment.

\begin{figure}[t]
    \centering
    \vspace{3mm}
    \includegraphics[width =80mm]{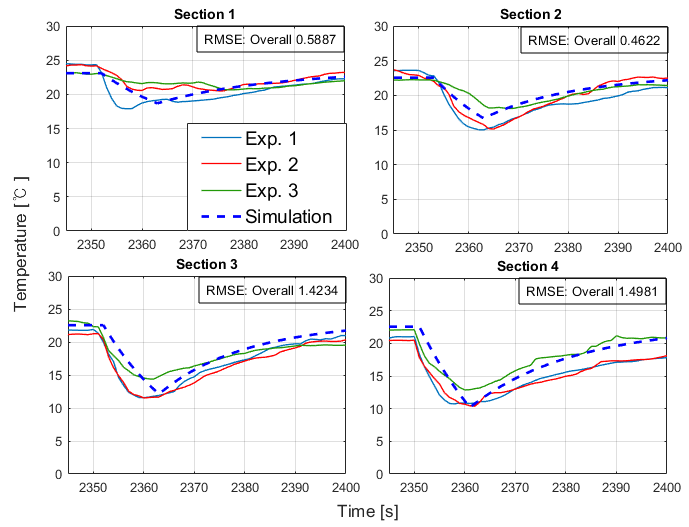}
    \caption{Assessing model coherence by comparing our simulation model with experiment results.}
    \label{validation result}
\end{figure}

In Fig.~\ref{validation result}, the comparison results are shown for three experiment iterations, visually distinguished by blue, red, and green lines. Despite consistent conditions, little variations in starting temperatures were observed across sections in the experimental data. 

Sections 3 and 4 exhibit significant temperature drops, with the impact of door-opening in Section 4 notably affecting Section 3 more than Sections 1 and 2. This difference may be attributed to seats hindering cold air propagation between front and back sections. Simulation results align with real-world trends, although it does not consider the air flow complexities. The root-mean-square errors between average experimental data and simulations for these sections are 0.5887, 0.4622, 1.4234, and 1.4981, affirming the model's accuracy in illustrating thermal interactions.

\section{NMPC-BASED HIERARCHICAL CONTROLLER DESIGN}

\begin{figure}[t]
    \centering
    \vspace{2mm}
    \includegraphics[width =80mm]{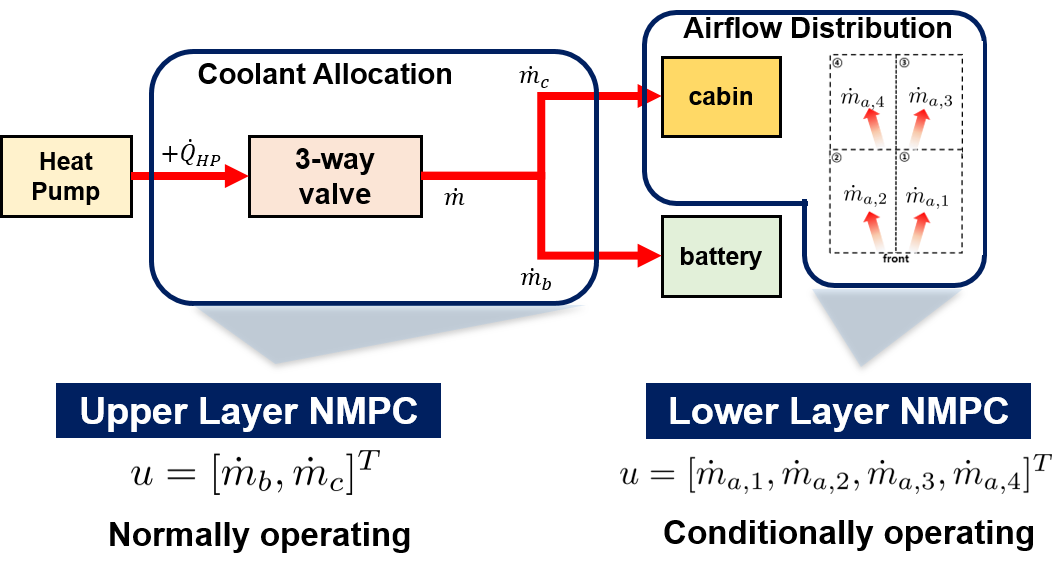}
    \caption{Hierarchical climate control system.}
    \label{Hier}
\end{figure}

In this section, we present the NMPC-based hierarchical ITMS heating control strategy, which consists of two layers as illustrated in Fig.~\ref{Hier}. The upper layer controller allocates the coolant to the cabin and battery sides, while the lower layer controller manages the distribution of inflow air to each section.

The upper layer primarily focuses on distributing the coolant to the cabin and battery sides. In this scenario, raising the battery temperature becomes crucial for improving efficiency, especially during vehicle starts in cold climates. This layer operates effectively when there are no interruptions, aiming to elevate the temperature to the target.

However, in situations like door-opening events, interruptions lead to varying temperature statuses in each section, creating discrepancies in passengers' thermal comfort. To address this, we introduce a hierarchical structure capable of managing inflow air distribution to minimize these discrepancies. The controller presented here lays the groundwork for further development in our upcoming studies.
\begin{align}
\label{MPC_formulation1}
\min_{u\in U} \ & \sum_{k=1}^{N_{p}} \alpha (T_{set,c}-T_{cab}(k))^2 + (1-\alpha) (T_{set,b}-T_{bat}(k))^2 \nonumber   \\
& \text{subject to} \nonumber \\
& \mathcal{X}_{H}(0)=\mathcal{X}_{H,0} \nonumber \\ 
& \mathcal{X}_{H}(k+1)=f(\mathcal{X}_{H}(k),u(k)) \nonumber \\
& \mathcal{X}_{H}=[T_{c1},T_{c2},T_{c3},T_{c4},T_{ha},T_{cab},T_{cb},T_{bat},SOC]^{T} \nonumber \\
& u = [\dot{m}_{b}, \dot{m}_{c}]^T \nonumber \\
& T_{cab}^{min} \leq T_{cab}(k) \leq T_{cab}^{max} \nonumber \\
& T_{bat}^{min} \leq T_{bat}(k) \leq T_{bat}^{max} \nonumber \\
& u_{min} \leq u(k) \leq u_{max} \  \nonumber \\
& \Delta u_{min} \leq \Delta u(k) \leq \Delta u_{max} \nonumber \\
& \Delta u(k) = u(k)-u(k-1) \nonumber \\
& k = 1,2,\cdots N_{p}  
\end{align}
 The cost function comprises two terms: the first term focuses on tracking the set temperature of the cabin with a corresponding weight factor $\alpha$, while the second term addresses the tracking of the battery set temperature with a weight factor $(1-\alpha)$. This controller efficiently manages the division of coolant flow, aiming to reduce energy consumption by adequately heating the battery temperature, which enhances the energy efficiency. To effectively regulate the cabin and battery temperatures, we establish maximum and minimum constraints. Additionally, for the development of a rational controller, constraints are imposed on the control inputs. To address the optimal distribution of the inflow air to each section, the second-layer MPC is formulated as follows:
\begin{align}
\label{MPC_formulation2}
\min_{u\in U} \qquad & \sum_{k=1}^{N_{p}} \alpha_{1:4}(T_{sec,1:4}(k)-T_{set,c})^2  +\beta(\Delta u(k))^2 \nonumber \\
& \text{subject to} \nonumber \\
& \mathcal{X}_{L}(0)=\mathcal{X}_{L,0} \nonumber \\
& \mathcal{X}_{L}(k+1)=f_{\text{open}}(\mathcal{X}_{L}(k),u(k)) \nonumber \\
& \mathcal{X}_{L}=[T_{s1},T_{s2},T_{s3},T_{s4},T_{ha},T_{cb}]^{T} \nonumber \\
& u = [\dot{m}_{a,1}, \dot{m}_{a,2},\dot{m}_{a,3}, \dot{m}_{a,4}]^{T} \nonumber \\
&  \mathcal{X}_{L,min} \leq \mathcal{X}_{L}(k) \leq \mathcal{X}_{L,max} \nonumber \\
& u_{min} \leq u(k) \leq u_{max} \nonumber \\
& \Delta u_{min} \leq \Delta u(k) \leq \Delta u_{max} \nonumber \\
& k = 1,2,\cdots N_{p}
\end{align}
 the formulation comprises two distinct terms in the cost function: the first term focuses on tracking the set temperature for each section, while the second term aims to prevent sudden changes in air flow for passenger thermal comfort. Excessive air distribution to a single section can adversely impact passenger thermal comfort, making it an impractical control action. To prevent this and ensure reasonable control, we impose constraints on the control inputs. Additionally, since this controller operates conditionally based on door-opening signals, we set state constraints to effectively prevent substantial temperature drops in specific sections.
Combining these two layers of controllers not only ensures effective management of the cabin temperature but also facilitates a more reasoned approach. This synergistic control system is designed to enhance thermal comfort for passengers, providing a balanced and efficient response to dynamic conditions, such as door-opening events. The integration of these layers creates a comprehensive control framework capable of addressing diverse scenarios and contributing to overall system stability.

\section{CASE STUDY}

In this section, we present an overview of the results and compare them with different control strategies. We conducted simulations and compared the cabin temperature outcomes for segmented sections utilizing three controllers: the proposed hierarchical controller, a single layer MPC, and a rule-based controller. The single layer MPC controller refers to a controller that solely comprises the upper layer MPC in Fig.~\ref{Hier} and remains fixed on air distribution even when door-opening occurs. This suggests that the results of the single layer MPC precisely match those of the hierarchical controller until the occurrence of door-opening. A comparison between the proposed controller and single layer controller will highlight the necessity of a hierarchical control strategy.

\begin{figure}[t]
    \centering
    \vspace{3mm}
    \includegraphics[width =80mm]{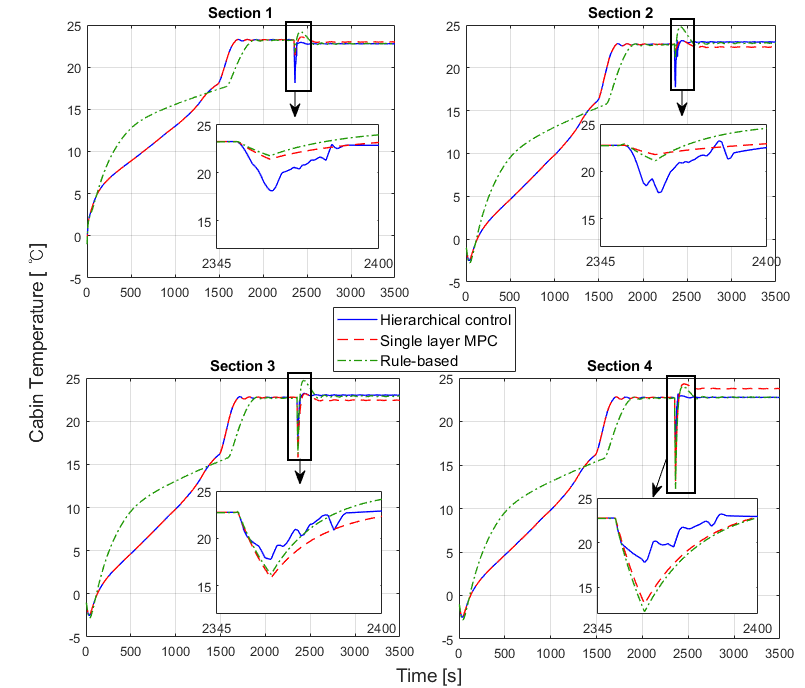}
    \caption{Comparison of temperature across cabin sections, including individual sections.}
    \label{whole_compare_results}
\end{figure}

The rule-based controller adheres to the following rules:
before the door opens, the heating process primarily focuses on warming the battery for temperature efficiency. The heating status is set according to Eq.~(\ref{mdotrule}).
\begin{align}
    & \dot{m}_{b} > \dot{m}_{c} \qquad (\text{if} \  T_{bat} < T_{\text{bound, min}}) \nonumber \\
    & \dot{m}_{b} \leq \dot{m}_{c} \qquad (\text{if} \  T_{\text{bound, min}} \leq T_{bat}).
    \label{mdotrule}
\end{align}
When the condition $T_{bound} \leq T_{bat}$ is met, the controller sets $\dot{m}_{c}$ to higher values to focus on the cabin heating. Meanwhile, the air inflow rates $\dot{m}_{a,1:4}$ remain constant until a new passenger boards. While driving, if a new passenger is detected, each section is heated differently in response to this signal, providing additional heat to the passenger's specific section.

\begin{figure}[t]
    \centering
    \includegraphics[width =80mm]{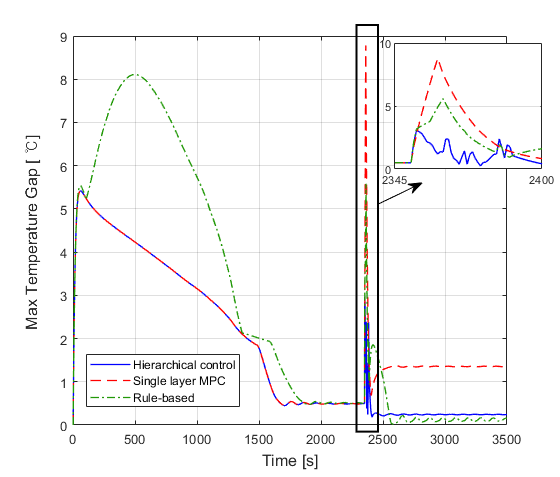}
    \caption{Maximum and minimum temperature gap between sections.}
    \label{MinMax}
\end{figure}

Figure~\ref{whole_compare_results} illustrates the heating simulation results for divided sections using the mentioned controllers. In the first and second graphs, the proposed strategy exhibits the most significant temperature drops, while the third and fourth sections show the smallest drop compared to others.
While allowing the temperature in the passenger's section to drop might seem unconventional, it is crucial to understand that our primary objective was to minimize the temperature gap between each section. This aspect can be expanded in our future work by integrating passengers' thermal comfort throughout the entire vehicle. We anticipate that incorporating a thermal comfort term into the cost function will eliminate this seemingly unconventional result.

In the graph depicting Section 4, where a significant temperature drop is anticipated, the proposed controller demonstrates superior performance. It exhibits a 46.96\% reduction in temperature drop compared to the single layer MPC and a 51.33\% reduction compared to the rule-based controller.

As depicted in Fig.~\ref{MinMax}, the proposed algorithm demonstrates the smallest temperature gap between sections during door-opening events. In contrast, the single layer MPC exhibits the largest gap during temperature recovery progress due to the fixed $\dot{m}_{a,1:4}$ throughout the entire simulation. In this graph, the proposed controller exhibits an 86.4\% reduction in temperature gaps compared to the single layer MPC and a 78.7\% reduction compared to the rule-based approach. This outcome underscores the significance of effective inflow air control and provides rationale for the development of the hierarchical multi-layered MPC controller presented in this paper. 
The reduction in temperature gaps between sections presents potential for further development, enabling effective management of multiple passengers' thermal comfort.

\begin{figure}[t]
    \centering
    \includegraphics[width =80mm]{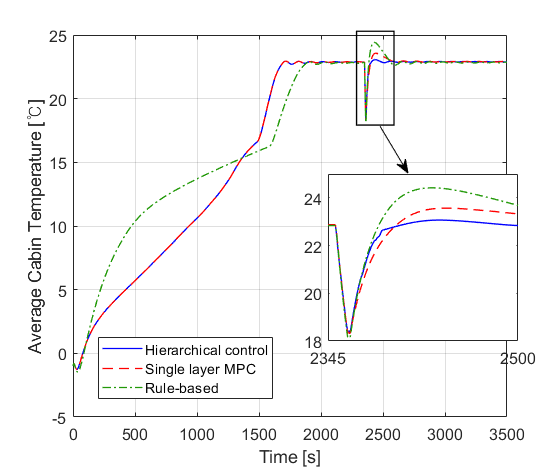}
    \caption{Comparison of Average Cabin Temperatures.}
    \label{평균온도}
\end{figure}

Figure~\ref{평균온도} shows the average cabin temperatures for each controller.
In this result, although the rule-based controller appears to return the temperature to the set point faster, the proposed control strategy exhibits the smallest overshoot, which can be a significant indicator of successful heating, while the rule-based one shows the biggest overshoot, and the single layer MPC follows closely. 
Numerically, the proposed controller achieves a 95.17\% reduction in temperature overshooting compared to the rule-based approach and an 87.9\% reduction compared to the single layer MPC. This indicates that the proposed hierarchical controller outperforms others in minimizing temperature overshoot and achieving swift recovery, particularly in the context of door-opening events.

\section{CONCLUSIONS}

In this study, we developed an integrated thermal management model for EVs, incorporating door-opening scenarios and validating the model through experiments to demonstrate its accuracy. The hierarchical control strategy proposed here showed efficient heating performance, particularly in minimizing temperature drops in the fourth section where the door-opening occurred. While seemingly conflicting results observed a greater temperature drop in the first section compared to other control strategies, this could be adjusted by assigning a larger weight factor $\alpha_{1}$ in Eq.~(\ref{MPC_formulation2}) to prioritize more heat for the driver section.
Although our primary focus was on minimizing temperature gaps across all sections, laying the groundwork for future investigations into the thermal comfort of multiple passengers within the vehicle cabin, we suggest that incorporating thermal comfort terms for each section into the cost function could eliminate significant temperature drops in passenger sections. This study establishes a rational research model for a heating strategy accounting for door-opening interruptions, with future endeavors aiming to optimize effective heating while considering individual thermal comfort for passengers in door-opening scenarios. Furthermore, with the foundation of V2X connectivity, there is significant potential for future research in advancing connected vehicle climate control. This aspect holds promising prospects for further exploration in the upcoming research endeavors.

\addtolength{\textheight}{-12cm}   



\bibliography{reference}    
\bibliographystyle{IEEEtran}

\end{document}